# Efficient local behavioral change strategies to reduce the spread of epidemics in networks

Yilei Bu[1], Steve Gregory[1], Harriet Mills[2]

[1] *Dept. of Computer Science, University of Bristol, Bristol BS8 1UB, UK*
[2] *Dept. of Infectious Disease Epidemiology, Imperial College, London W2 1PG, UK*
Email: steve@cs.bris.ac.uk

It has recently become established that the spread of infectious diseases between humans is affected not only by the pathogen itself but also by changes in behavior as the population becomes aware of the epidemic; for example, social distancing. It is also well known that community structure (the existence of relatively densely connected groups of vertices) in contact networks influences the spread of disease. We propose a set of local strategies for social distancing, based on community structure, that can be employed in the event of an epidemic to reduce the epidemic size. Unlike most social distancing methods, ours do not require individuals to know the disease state (infected or susceptible, etc.) of others, and we do not make the unrealistic assumption that the structure of the entire contact network is known. Instead, the recommended behavior change is based only on an individual's local view of the network. Each individual avoids contact with a fraction of his/her contacts, using knowledge of his/her local network to decide which contacts should be avoided. If the behavior change occurs only when an individual becomes ill or aware of the disease, these strategies can substantially reduce epidemic size with a relatively small cost, measured by the number of contacts avoided.

## I. INTRODUCTION

### A. Background

Network models have been widely used in epidemiology to study the spread of infectious diseases. In networks, a vertex represents an individual and an edge between two vertices represents a contact over which disease transmission may occur. An epidemic spreads through the network from infected to susceptible vertices.

Once an epidemic emerges, individuals tend to make behavioral changes to protect themselves or others; for example, by immunization or quarantine/isolation. The effects of such behavioral changes can be modelled in at least three ways [1]. Immunization can be modelled as a change in disease state where immunized vertices change from susceptible to immune so they cannot be infected or transmit disease [2-6]. Social distancing can be modelled by a modification of epidemic parameters, for example, reducing transmission rate, or a modification of the network structure, in which an existing edge is removed or rewired [7-10]. We focus on control strategies that implement edge removal, for which we can identify four issues:
1. Who is responsible for removing edges?
2. Which edges are removable?
3. How are edges chosen for removal, from among all removable edges?
4. When are edges removed?

The first issue concerns whether edges are removed by some central authority (for example, by quarantining individuals), or by the individuals themselves (by distancing themselves from their contacts). We call these global and local strategies, respectively. With a *global* strategy, any edge in the network is removable, and information about the entire network can, in principle, be used to choose edges for removal. An example is the targeting of high-degree individuals in a network [11,12]; this is called a local strategy in Ref. [12] because it can be computed using only local properties of vertices, but we consider it global because it relies on knowledge of all vertices in the network. Alternatively, high-betweenness vertices [5,11] or edges [7,8] are sometimes targeted in networks with community structure, to stop disease spreading between communities. However, the computation of betweenness has high time complexity. Recently, alternative immunization strategies have been proposed, which seek to fragment a network into small susceptible components [13,14]. Nevertheless, the drawback of all global strategies is that they require information about the whole network, which is unlikely to be available for real contact networks.

In contrast, with a *local* strategy, each vertex can only remove its own edges, and it chooses between them using only local information about its neighbors and possibly its neighbor's neighbors, etc. Local strategies correspond to the kinds of strategy that are feasible in real life when individuals change their behavior in response to an epidemic. Additionally, because they use only local information, they tend to be faster to compute. Existing local strategies mostly take the individual's fear of disease into consideration. For example, with the "SI link" strategy, a susceptible vertex will choose to remove edges connecting it with infected vertices [9,15,16]. This is very efficient because these are the only edges capable of spreading infection. However, it may be unrealistic because (a) susceptible individuals might not be aware in time that a contact is infected, (b) infected individuals might not know which contacts are susceptible, and (c) some diseases have an pre-symptomatic infectious phase, during which neither the infected nor the susceptible individual know that the edge joining them should be removed. Moreover, this strategy has little, if any, effect in the early stages of an epidemic because there are then few infected vertices.

Finally, we can consider *when* to remove edges: global control strategies are usually applied independently of the epidemic spread while local control strategies are applied during the epidemic. With local strategies, a vertex can remove (or rewire) edges [9,15,16] or be immunized [6] when infection is detected in a contact. Detecting an infectious contact can be hard in practice, and may be too late to prevent infection. One option is to take similar action when a vertex becomes *aware* of disease, for example, through propagation of information [17] or when detected nearby in the network [18,19].

### B. Our approach

In this paper we propose the use of local edge-removal strategies to prevent the spread of epidemics. Each vertex ranks its edges according to various measurements and the top-ranked fraction of edges are chosen for removal. We remove only a fraction of edges because an infected individual is likely to reduce his/her social contacts while infected but is unlikely to become completely isolated. This is also less costly than trying to remove or rewire all edges of an infected vertex.

We do not require individuals to know the disease state of any of their contacts. Instead, we assume that each individual knows the structure of his/her local network (the neighbors and all of the neighbors' edges). This seems practical because people are more likely to know how their contacts usually interact than to know the current state of their health.

We expect individuals to change their behavior, by removing edges, when they become infected and develop symptoms, rather than while trying to avoid infection. We believe this is realistic because people often become less mobile when ill and will seek to avoid spreading the infection. However, this cannot prevent the spread of diseases that have an initial pre-symptomatic infectious phase. To do this, we simulate awareness [17]: when a vertex enters the symptomatic infected state, it will not only become aware and apply its local control strategy but will also spread awareness through the network; aware vertices will also apply the control strategy.

Most existing local control strategies [9,15,16,18,20] have not taken community structure into consideration. However, all of our strategies are based on community structure, which has been detected in studies of real-world social networks [21] and is known to affect the spread of epidemics [5]. The strategies we propose all have the same aim: an individual should try to avoid contact with those in different communities. We exploit existing local community detection methods to determine which edges an individual should remove.

In the rest of the paper we experiment with several local strategies of this type, and compare them with some popular existing strategies. We assume a class of influenza-like infectious diseases which are transmitted from person to person by respiratory or close-contact means. The next section defines the epidemic model that we assume, types of behavior change, and then describes our local control strategies and the existing strategies which we use for comparison. Section III presents the results of our experiments on simulating disease spread in the presence of various strategies. Section IV discusses the results and draws some conclusions.

## II. METHODS

### A. Networks

We create multiple networks characterizing populations: individuals are represented by vertices and their contacts are represented by edges. We assume a static network, a reasonable assumption for a short-duration, fast spreading infection. We use the following network types:

*Random* (1000 vertices, ~5000 edges) is a set of artificial networks generated using the Erdös-Rényi binomial model. Each pair of vertices is connected with probability 0.01.

*Exponential* (1000 vertices, ~5000 edges) is a set of artificial networks with each pair of vertices connected with probability 0.01. They are then rewired to an exponential degree distribution using a greedy rewiring algorithm [22].

*Scale-free* (1000 vertices, ~5000 edges) is a set of artificial networks generated using an improved preferential attachment method [23]. The exponent is $\gamma=1.6$. Each time a vertex is added, edges are created between it and existing vertices with probability proportional to existing vertices' degrees.

*LFR* (10000 vertices, ~80000 edges) is a set of artificial networks with community structure and a power-law degree distribution, constructed by the "benchmark" algorithm of Lancichinetti, Fortunato, and Radicchi (LFR) [24]. The networks have average (maximum) degree of 16 (32). The mixing parameter, which controls the community strength, is 0.1. Community sizes range from 50 to 100.

*Blogs* (3982 vertices, 6803 edges) is a network of blogs on the Windows Live Spaces platform [25]. This network has strong community structure.

*PGP* (10680 vertices, 24316 edges) is a "web of trust" based on use of the PGP algorithm in July 2001 [26].

*Netscience* (379 vertices, 914 edges) is a small collaboration network [27].

*School* (657 vertices, 1139 edges) is generated from a dataset collected using wireless sensor network technology during a typical day at an American high school [28]. It records close-proximity interactions, and their duration, between students, teachers, and staff. This dataset is particularly relevant to influenza-like diseases, so it is highly suitable for our purposes. We only retain contacts with duration longer than 50 minutes.

### B. Epidemic model

We simulate the spread of an influenza-like virus transmitted by respiratory or close-contact means over the network of contacts. We use a SPIR model, which is like the familiar SIR (Susceptible-Infected-Recovered) model but infectious in both the pre-symptomatic (P) and symptomatic (I) stages. Initially all vertices are in state S and one vertex is chosen randomly to be in state P. At each time step (one day), a vertex in state S can be infected by a neighbor in state P or I with probability $\beta$. Therefore, for a vertex with $k$ infectious neighbors, the probability of becoming infected will be $1-(1-\beta)^k$. Once infected, a vertex moves from state S to state P. Each vertex in state P takes $p$ time steps to change to state I and then enters the recovered state, R, with probability $\gamma$. Over time the infection spreads through the network; the simulation is halted once no vertex is in state P or I. We do not include birth or death in the model, because we assume a relatively short, non-fatal infection. When $p=0$, state P is removed and the model becomes the traditional SIR model.

$R_0$ (the basic reproduction number) represents the average number of new infections caused by one infectious individual in an entirely susceptible population during its infectious period. If $R_0<1$ the infection will die out, otherwise infection will spread through the population.

In a network with heterogeneous degree distribution, $R_0$ depends on the degree distribution according to the equation $R_0 = \rho_0 \langle k^2 \rangle / \langle k \rangle^2$, where $\langle k \rangle$ is the mean degree, $\langle k^2 \rangle$ is the mean squared degree. $\rho_0$ (a lower bound on $R_0$) is defined as $\beta \langle k \rangle / \gamma$, where $\beta$ is the transmission probability for the disease and $\gamma$ is the recovery rate [29]. We fix $\gamma=0.2$ (the inverse of the infectious period of 5 days, which includes both P and I phases) and $\rho_0$ to 3, and set $\beta$ for each network according to $\beta = \rho_0 \gamma / \langle k \rangle = 0.6/\langle k \rangle$. This results in a different value of $R_0$ for each of our networks (e.g., 12.1 for Blogs, 12.4 for PGP, 5.0 for Netscience, 4.2 for School), but it is fixed for each network, allowing a fair comparison of strategies.

The range of $R_0$ values that we use is plausible for influenza and other respiratory diseases, such as SARS [30]. These highly-contagious diseases have a large media presence, which raises awareness of the disease and potentially causes risk-averting behavior in the population, such as we model here.

### C. Behavior model

Funk et al. [1] classified all behavior models according to the source and type of information that result in people's behavioral changes. The source of information can be *global* (publicly available information such as newspapers, TV news, and other media) or *local* (taken from the social or spatial neighborhood only). The types of information can be classified as objective (*prevalence-based*: directly related to disease prevalence) and subjective (*belief-based*: having nothing to do with disease prevalence) [1]. We use the following classifications.

*Global belief-based model.* The whole population becomes aware of the disease before it actually starts spreading. All vertices in the network apply their local control strategy before timestep 0. We compare local and global strategies.

*Global prevalence-based model.* Awareness of the disease spreads globally, via the media, for example. The disease spreads unchecked at first. After a certain time, all vertices not yet infected will become aware and apply the local strategy. We vary the time of awareness occurring, to compare local control in the global belief-based and prevalence-based models.

*Local prevalence-based model.* An individual changes behavior only in response to local disease prevalence. Here, vertices apply their control strategy only when they enter a (symptomatic) infectious state. We use this model to compare our proposed local strategies with existing ones.

*Local belief-based model.* If a local strategy is applied only when symptoms appear it may be too late to prevent infected vertices from infecting their contacts. In contrast, in the local belief-based model, asymptomatic infected vertices (in state P) and susceptible vertices (S) may become aware and apply their local control strategy. Awareness originates with symptomatic infected vertices and spreads to neighbors with a specific probability, which controls the speed of awareness spread. Both the probability of awareness spread and the duration of the presymptomatic phase can be varied.

### D. Local control strategies

Our local control strategies are all executed for each vertex and remove up to a certain fraction of that vertex's edges. The parameter *remove fraction* specifies an upper limit on the number of edges that each vertex can remove. Each vertex will use the strategy repeatedly to find the best candidate edge to remove. If the upper limit is reached, no more edges will be removed from that vertex. Conversely, if a vertex cannot find any candidate edge, even if the upper limit has not been reached, the vertex will not remove any further edges. This could happen either because the strategy does not consider the remaining edges worth removing or because the remaining edges lead to vertices with no other neighbors. We never remove an edge to a vertex with degree one even if this edge is the best choice calculated by the strategy (so that no individual becomes totally isolated).

We have implemented various local strategies. The first four are based on community structure, while the fifth uses vertex degree. The computation of the strategies is illustrated in Fig. 1. We compare them with a *random* strategy, in which we choose an edge to delete at random from among the vertex's edges.

*Similarity strategy.* The Jaccard similarity coefficient [31] measures the ratio of common neighbors between two vertices, $i$ and $j$:

$$\sigma_{ij} = \frac{|N_i \cap N_j|}{|N_i \cup N_j|}. \tag{1}$$

A low coefficient suggests that the two vertices may belong to different parts of the network structure (e.g., different communities) and the edge connecting them will become a good target to remove, resulting in separating the two communities. The method we use [7,8], for vertex $i$, is to calculate the similarity $\sigma_{ij}$ between $i$ and each of its neighbors $j$, and then remove the edge leading to the neighbor that has the lowest similarity.

*Clustering coefficient (CC) strategy.* The clustering coefficient of a vertex $i$ is the fraction of pairs of neighbors of $v$ that are connected by an edge [10]. It measures how closely its neighbors are connected to each other, which is related to the extent to which $i$ and its neighbors belong to the same community. Our strategy works, for vertex $i$, by first calculating the clustering coefficient of $i$ and then calculating the clustering coefficient of $i$ after provisionally removing each of $i$'s edges separately. The edge chosen for removal is the one that can increase the clustering coefficient most.

*Local fitness maximization (LFM) strategy.* The fitness of the subgraph comprising a vertex and its neighbors is calculated using the number of internal and outgoing edges for the subgraph, as defined in [32]; it gives an indication of how community like the subgraph is. Our strategy treats vertex $i$ and its neighbors as a subgraph $g$ and calculates the subgraph fitness $f_g$:

$$f_g = \frac{k_{in}^g}{(k_{in}^g + k_{out}^g)^\alpha}, \quad (2)$$

where $k_{in}^g$ and $k_{out}^g$ are the total internal and external degrees of the vertices in community $g$, and $\alpha$ is a positive parameter that controls the size of the community (we use $\alpha=2$ for all of our simulations). It then, for each neighbor $j$ of $i$, calculates the fitness for subgraph $g$ without $i$. Finally, it removes the edge with the maximum fitness value; if no edge has a positive fitness value, no edge will be removed.

*Local modularity (LM) strategy.* Clauset's local community detection algorithm [33] defines a measure of local community $C$ by considering its boundary $B$ (a subset of $C$). $B$ contains those vertices in $C$ that have at least one neighbor in $U$, the unknown part of the network. Local modularity [33] is defined as:

$$R = \frac{\sum_{ij} B_{ij} \delta(i,j)}{\sum_{ij} B_{ij}} = \frac{I}{T}, \quad (3)$$

where $B_{ij}$ is 1 if vertices $i$ and $j$ are connected and either vertex is in $B$, or 0 otherwise. $\delta(i,j)$ is 1 when either $i \in B$ and $j \in C$ or vice versa, or 0 otherwise. $T$ is the number of edges that have at least one endvertex in $B$ and $I$ is the number of edges with one endvertex in $B$ and the other in $C$.

Our strategy treats vertex $i$ and its neighbors as the local community $C$, finds the corresponding $B$, and calculates $R$. Then, for each neighbor (of $i$) $j$ in $B$, it provisionally removes $j$ from $C$, finds the new $B$ and calculates the new $R$. Finally, it chooses the vertex $j$ such that $R$ is maximally increased, and removes edge $\{i,j\}$. If no edge removal increases $R$, none will be removed.

*High degree (HD) strategy.* High-degree vertices are more likely to be infected and more likely to spread infection to others, so they are often targeted to reduce disease spread. Unlike our other local strategies, this one is independent of community structure. Our strategy, for vertex $i$, simply calculates the degree of all neighbors of $i$ and removes the edge to the one with the highest degree.

### E. Strategies used for contrast

Here we describe several alternative strategies which we have used for comparison with our proposed local strategies.

*Global edge betweenness (GEB) strategy.* The betweenness of an edge is defined as the number of shortest paths between all pairs of vertices that pass along the edge. Edges with high betweenness tend to have a more important role in delaying or preventing the disease spreading [11]. Our GEB strategy calculates the edge betweenness of the network and removes the edge with the highest betweenness that does not have an endvertex with degree one. This is repeated until $rm$ edges have been removed, where $r$ is the remove fraction and $m$ is the number of edges in the network. Because this strategy is computationally expensive (time $O(rm^2n)$), we only use it in small networks. We compute edge betweenness using the method of Ref. [34] but do not use its community detection algorithm.

*Infomap strategy.* Infomap [35] is a widely-used community detection algorithm. Our strategy is to use Infomap to find all communities and hence all intercommunity edges. We then randomly choose a specified number ($rm$) of these intercommunity edges to remove, never choosing an edge to a vertex with degree one. Like edge betweenness, the Infomap strategy is global because it is computed from information about the whole network, but it is much faster (time $O(m)$ [36]), so we can use it on large networks.

Although our edge-removal strategies (Similarity, CC, LFM, LM, and HD) use only local information, they can be compared with these two global strategies when all vertices in the network apply local edge-removal strategies at the same time. For global strategies, the parameter *remove fraction* specifies the fraction of the whole network's edges that are removed, while for local edge-removal strategies, it specifies the fraction of the edges of each vertex. In both cases, these are upper bounds on the fraction of the whole network's edges that actually are removed. We refer to the total fraction of edges removed by the term *total remove*.

*SI-link strategy.* In the SI-link strategy, susceptible vertices remove their edges to infected vertices [1,15,16,37]. This strategy is executed by vertex $v$ when it becomes infected. It first finds $S(v)$, the set of its susceptible neighbors with degree greater than 1 and then removes edges to all susceptible neighbors up to an upper limit of $e=\lfloor kr \rfloor$, where $k$ is the degree of $v$ and $r$ is the remove fraction. If $|S(v)| > e$, $e$ vertices are randomly selected from $S(v)$ and only edges to these are removed.

## III. RESULTS

Each strategy is a trade-off between disease control (the benefit) and the number of edges removed (the cost); therefore we measure both aspects. The effectiveness of disease control is assessed by measuring the total epidemic size: the fraction of network vertices that become infected and recover during the epidemic. All results are averages of 500 simulations (each using a different network instance, in the case of the artificial networks). We include simulations that result in no epidemic, because it is impossible to distinguish between those that occur by chance and those that are due to the success of the strategy.

### A. Global belief-based model

In this section we use the SIR model and apply each strategy before the epidemic begins.

In a random (Erdös-Rényi) network (Fig. 2(a)), the epidemic size decreases when the remove fraction (the upper limit on the fraction of their edges that each vertex can remove) increases, and the epidemic dies out when the remove fraction reaches 50%. There exists a threshold (20% remove fraction), below which Clustering Coefficient (CC) and Similarity give better disease control than the random strategy. Beyond this point, the Local Fitness Maximization (LFM), Local Modularity (LM), and High Degree (HD) strategies perform better than random. Figure 2(b) shows that HD and LM remove slightly fewer edges than the random and LFM strategies, because they have stricter requirements when choosing candidate edges to remove. Therefore, the HD and LM strategies are best overall since they remove generally fewer edges than the random strategy but result in reduced epidemic size. However, there is very little difference between all local strategies and random edge removal.

Figure 2(c) combines Figs. 2(a,b) by showing how epidemic size varies with the fraction of edges actually removed ("total remove"). In most of our experiments we show the results in this compact form only.

The same situation is also revealed in the exponential and scale-free networks (Fig. 3), and the difference between strategies is more striking. The remove fraction threshold in both networks is over 25%. Beyond this threshold, HD and LM cause the smallest epidemic with the fewest total removed edges, of all local edge-removal strategies.

In the LFR network (Fig. 4(a)), the epidemic size also decreases when the remove fraction increases, but the LFM, LM, Similarity, and CC strategies cause the epidemic size to decrease much more rapidly than the HD and random strategies for the fraction of edges removed. The LFM strategy is slightly less effective than LM, Similarity, and CC, but all four strategies reduce the epidemic size to about 10% with a remove fraction of 20%. The LM strategy removes fewer edges than the other local strategies, making it the best local strategy for the LFR network. An additional benefit of LM is that the strategy will not continue to remove many more edges after all critical (intercommunity) edges have been found.

We would expect our strategies to work better in networks with stronger community structure. To test this, we use the same LFR network as Fig. 4(a) but increase the mixing parameter from 0.1 to 0.3 (Fig. 4(b)) and 0.5 (Fig. 4(c)). This shows that our strategies (Similarity, CC, LM, LFM) all perform worse as mixing increases, because the community structure effectively disappears as mixing increases.

In our real networks, all local edge-removal strategies perform better than random removal, resulting in a lower epidemic size (Fig. 5). Overall, LM is the most effective method.

In order to compare local edge-removal strategies with global strategies, we used the Infomap and Global Edge Betweenness (GEB) strategies. For these strategies, *remove fraction* means the fraction of the whole network's edges that are removed, while for local edge-removal strategies, it refers to the upper limit for each vertex, as explained in the Methods.

When the total number of removed edges is the same, the performance of Infomap is very similar to that of the local edge-removal strategies in the random, exponential, scale-free, and LFR networks (Figs. 2, 3, 4(a)). For example, in the LFR network (Fig. 4(a)), the Infomap strategy also results in a very small epidemic at only 10% remove fraction, and it never removes more than 10% of the network's edges. This is because Infomap successfully removes all intercommunity edges, stopping the disease spread. The same occurs in the Blogs, PGP, and Netscience networks (Fig. 5) since these also contain strong community structure. Although the local strategies cannot perform as well as Infomap in these networks, the best (Similarity and LM) reduce the epidemic to around 15% of the network size after removing the same number of edges as the Infomap strategy. In the School network (Fig. 5(c)), the local edge-removal strategies perform slightly better than Infomap.

The GEB strategy results in a much lower epidemic size (less than half) compared with other strategies at low remove fraction. However, once all intercommunity edges have been found, GEB (unlike Infomap) continues to remove more edges as the remove fraction increases. Recall that GEB is computationally intensive so we test this strategy only on the smallest real networks (Fig. 5(c,d)).

For clarity, we have removed error bars from our plots of epidemic size, leaving one illustrative example, the 10000-vertex LFR network. Figure 6 shows how the epidemic size varies with the remove fraction, including error bars. The variation in epidemic size is quite high; this is caused by the small number of epidemics which never take off in the population (irrespective of the success of the strategy). In contrast, the fraction of edges removed is very stable.

Examining the average epidemic size does not reveal whether the control strategies act to suppress the size of all epidemics or reduce the probability of full-scale outbreaks occurring. Therefore, in Fig. 7 we explore this for the same example as Fig. 4(a): the 10000-vertex LFR network with mixing parameter 0.1. Figure 7(a) shows the fraction of epidemics whose size exceeds 2% of the network size, which is the same threshold as used in Ref. [5], while Fig. 7(b) plots the average size of these large outbreaks only. This demonstrates that the strategies both reduce the chance of an epidemic taking off and reduce the size of any outbreak that does occur.

### B. Global prevalence-based model

So far, we have assumed that the whole population becomes aware, and changes its behavior, before the epidemic starts to spread. In reality, there may be a delay before the population becomes aware of the disease, during which the disease prevalence increases. Figure 8 shows how the prevalence increases with *global step*, the number of time steps since the start of the epidemic. To measure the effect of delay, we assume that all members of the population become aware and change their behavior, simultaneously, at some time after the epidemic starts to spread.

In Fig. 9 we use one artificial network (LFR) and one real-world network (Blogs) that has community structure, since we found that local edge-removal strategies perform best in networks with community structure. In the plot, *global step* means the number of time steps between the start of the epidemic and the application of the strategy. A global step of 0 means that the strategy is applied before the epidemic begins.

In both networks, if the strategies are applied before the disease starts, Infomap and some local strategies can keep the epidemic size below 10%. With a delay in applying the strategies, the epidemic size grows steadily for all strategies and, when the delay is large enough, effective strategies tend to perform as badly as random removal. However, the relative merit of the strategies is preserved. This suggests that a good strategy is still worth using even if it cannot be applied early. The number of edges removed is not affected by the delay.

### C. Local prevalence-based model

Next we evaluate our local prevalence-based model, in which each vertex applies its strategy only when it becomes infected instead of the whole population acting together. We compare it with the global belief-based model, in which the strategy is applied before the epidemic begins. Again, we use the LFR and Blogs network to illustrate the strategies (see Fig. 10).

The results show that when the remove fraction is small performance is improved when the strategies are applied early (i.e., global) compared to when they are applied in the presence of infection (i.e., local). However, as the remove fraction increases, this superiority diminishes until finally they exhibit similar performance. Importantly, in the local prevalence based model the number of edges removed remains small and decreases rapidly as the remove fraction increases. This is because the disease dies out so quickly that only a few vertices need to remove edges. In contrast, with the global belief-based model, all vertices remove edges independently of infection, so the number of removed edges keeps increasing. For example, in the LFR network (Fig. 10(a,b)), LM(local) and LM(global) have a similar low epidemic size at 15% remove fraction, but LM(local) removes a negligible number of edges while LM(global) removes about 10%.

Therefore, if the remove fraction is high enough, the local prevalence-based model can be as effective as the global belief-based model, with far fewer edges removed. This indicates that the best method of disease control is for only infected individuals to make behavioral changes, provided that they remove enough of the most critical edges before infecting others; for a severe flu-like illness, this is potentially quite likely. This way, far fewer edges need to be removed and the potential economic cost is much lower.

In Fig. 11 we compare our local edge-removal strategies with the SI-link strategy, which mimics the human instinct to reduce contact with infected people and can only be applied if vertices' disease states are visible to their neighbors. At 10% remove fraction in the LFR network (Fig. 11(a,b)), the epidemic size with SI-link is similar to other local edge-removal strategies, and they remove the same number of edges. But as the remove fraction increases, all local edge-removal strategies perform better than SI-link and also remove fewer edges. At 10% remove fraction, CC, LFM, LM, and Similarity result in a negligible epidemic size with a negligible number of edges removed, while SI-link has very little impact on the epidemic and 10% of edges removed. A similar phenomenon is seen in the Blogs network (Fig. 11(c,d)). Although the advantages of local edge-removal strategies, compared with SI-link, are not as obvious as in the LFR network, local edge-removal strategies still have better epidemic control and remove fewer edges.

### D. Local belief-based model

We now consider a disease with a presymptomatic but infectious phase (SPIR model). We allow disease awareness to spread through the network simultaneously with disease spreading (and between the same contacts); we spread awareness at a certain rate, which can vary. Since our local edge-removal strategies are independent of disease state, an individual can take action as soon as it becomes aware of disease nearby, rather than waiting until its own symptoms develop, which may be too late. Therefore, a vertex becomes aware when it becomes symptomatic *or* when it receives awareness from its neighbors. We use the LM edge-removal strategy throughout. The total time infected (presymptomatic plus symptomatic) is five days.

Increasing the duration of the presymptomatic infectious phase ($p$) increases the epidemic size (Fig. 12). The faster the awareness spreads ($w$), the lower the epidemic size and the more edges are removed. In general, the spread of awareness offsets the effect of a longer presymptomatic infectious phase. In the LFR network, with $w=0.3$ and $p\leq4$, the epidemic is smaller than with $p=1$ and no awareness spread. In the Blogs network, $w=0.3$ achieves this result for $p\leq2$ and has no effect for $p\geq3$.

## IV. DISCUSSION

In networks with community structure, we have shown that edge-removal strategies which reduce intercommunity contact based on local information can reduce epidemic size. In many of the networks we consider, only a small fraction of edges need be removed to result in a significant drop in epidemic size.

Using the global belief-based model, the community-based strategies (LM, LFM, Similarity, CC, Infomap, and GEB) all perform significantly better than random, and generally outperform HD. Of the local strategies (LM, LFM, Similarity, and CC), LM stands out as generally the most effective, and it tends to remove fewer edges than the other strategies. While Infomap and GEB perform well, they are global strategies and have very high time complexity. Because complete network information is rarely available in real life, LM is more practical to use than Infomap and (especially) GEB.

The local strategies implement different ways to minimize contacts between communities. For example, for the LM and LFM strategies, individuals essentially remove links to those contacts that have most contacts external to the community. In a similar way, the CC and Similarity strategies measure the strength of the community an individual is in, and improve the measure by removing links which bridge communities. In reality, these strategies mean reducing contact with individuals who travel frequently, or who work in a high contact job (teaching is a good example). Isolated communities can protect themselves (have less chance of becoming infected) and protect others (prevent an infection being transmitted onward).

We obtained similar results when strategies were applied only after a high global prevalence was reached. Even though the strategies all deteriorate when applied later, their relative performance is unchanged. This kind of delay is very likely to occur in practice, because of the time needed to detect the epidemic and implement the strategies.

Experiments on the local prevalence-based model show that, for relatively high remove fraction, applying each strategy when infected can be almost as effective as applying it in advance, while removing far fewer edges. Again, local strategies (LM, LFM, Similarity, and CC) are more effective than the traditional SI-link strategy, despite removing fewer edges. This is because they focus on intercommunity edges rather than randomly cutting edges between infected and susceptible vertices. Additionally, these strategies are practical as they do not depend on individuals knowing the disease states of others.

The basic LM strategy also performs well in the local belief-based model, when infected vertices become aware immediately after being infected. If vertices apply edge-removal strategies before infecting other vertices, there is no need to spread awareness. However, when there is a presymptomatic infectious phase, a higher speed of awareness spread can help to reduce the epidemic size. The effect is reduced if the presymptomatic infectious phase is too long. Interestingly, awareness can keep the epidemic size as small as when the whole population applies the strategy before the disease starts, but with far fewer removed edges. This corresponds to people becoming aware of disease in their local community instead of through mass media [19].

In our behavior model, we assume that removed edges remain broken for the duration of the epidemic. It is plausible that individuals might reinstate removed edges after a certain length of time or after they cease to be infected (if they ever were) [9]. In a future project, it would be worth experimenting with this model, and possibly one in which edges are rewired [15] instead of removed.

It would also be of interest to vary the values of $\beta$, $\gamma$, $p$ and the *remove fraction*. We anticipate that altering these parameters could impact the effectiveness of the strategies. Similarly, introducing heterogeneity across the network, for example giving individuals different remove fractions, susceptibility or infectiousness, will also impact the efficacy of strategies.

Among our local strategies, the one that works best is based on Clauset's local modularity [33]. However, improved variants of this function (e.g., [38]) have been proposed more recently, and these should also be evaluated in future work.

We found little difference between our strategies on artificial networks without community structure. Community structure is a common feature of contact networks: mixing between individuals can cause distinct communities, such as households, work, and school groups.

Our strategies work best when the community structure is disjoint. However, contact networks often contain communities that overlap [25,39]. Even if the overlapping communities can be found, they cannot be separated simply by removing a few edges. Our solution is to consider the contact networks as *weighted*: the weight of an edge represents the transmission rate, $\beta$. Then the communities formed by higher-weight edges are more likely to be disjoint than the communities containing all edges. For example, assuming every individual is a member of one "home" community and one "work" community, these communities appear inextricably linked. However, if the "home" edges have a higher transmission rate, the communities become effectively disjoint: the "work" edges can be disregarded (and removed), corresponding to a simple "stay at home" strategy. To handle overlapping communities in general, we will need to estimate the transmission rates on contact network edges, and extend our local control strategies to weighted versions.

Finally, each edge-removal strategy is a trade-off between the beneficial effect of reducing the epidemic and the economic cost of avoiding contacts. If we could quantify these costs and benefits in the same units, we could directly calculate the net economic benefit of applying each strategy, as done in Ref. [18].

We have proposed a new approach to epidemic control that we believe is both practical and effective. It is practical because there is no assumption that individuals or central authorities know the structure of the whole network, which could comprise billions of people, and no need for individuals to know the disease state of their contacts.

Instead, individuals effectively run a local community detection algorithm on their local network to determine which contacts to avoid. We have shown that this is effective even if individuals wait until they become infected and contagious before changing their behavior. The same strategies also work (potentially better) if applied earlier: for example when individuals become aware from neighbors or through mass media. These scenarios are realistic; several studies [40-43] have shown that health-related behavior of individuals is often influenced by their social contacts.

As well as being more realistic than global strategies, these local strategies can achieve (almost) comparable results in controlling an epidemic and they tend to be far less costly in terms of the number of edges removed.

All of our successful strategies perform a limited, local, form of community detection. As such, they can be seen as an important application area for community detection algorithms [36]. Nevertheless, the strategies are simple enough that they can easily be executed by individuals, with limited knowledge of their contacts, as soon as an epidemic emerges.

## ACKNOWLEDGEMENTS

We are grateful to the anonymous referees for their helpful and perceptive comments.

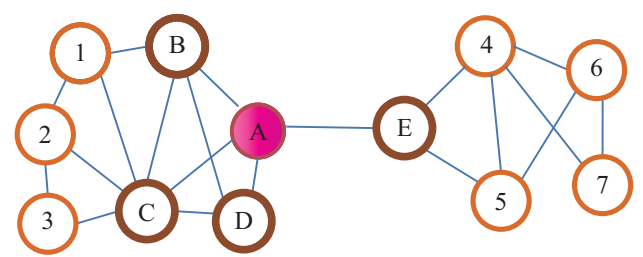

| Strategy | Chosen edge | Calculation of score |
|---|---|---|
| Similarity | $e_{AE}$ | $\sigma_{AE} = 0$ |
| CC | $e_{AE}$ | $\Delta CC = \dfrac{3 \cdot 2}{3 \cdot 2} - \dfrac{2 \cdot 3}{4 \cdot 3} = 0.5$ |
| LFM | $e_{AC}$ | $\Delta f_C^g = \dfrac{2 \cdot 4}{(2 \cdot 4 + 5)^2} - \dfrac{2 \cdot 7}{(2 \cdot 7 + 6)^2} = 0.012$ |
| LM | $e_{AE}$ | $\Delta R_E = \dfrac{6 - 0 - 1 + 0 - 0}{12 - 2 - 1 + 0 - 0} - \dfrac{6}{12} = 0.056$ |
| HD | $e_{AC}$ | Vertex C has the largest degree (6). |
| GEB | $e_{AE}$ | Edge $e_{AE}$ has the highest value of edge betweenness (35). |
| Infomap | $e_{AE}$ | Two communities are found: {4,5,6,7,E} and {1,2,3,A,B,C,D}, $e_{AE}$ is the only intercommunity edge. |

FIG. 1. (Color online) A simple network, and calculation of edge scores for edge-removal strategies for vertex A. The local strategies (Similarity, CC, LFM, LM, and HD) use only information about A and its neighbors (the vertices drawn with thick circles).

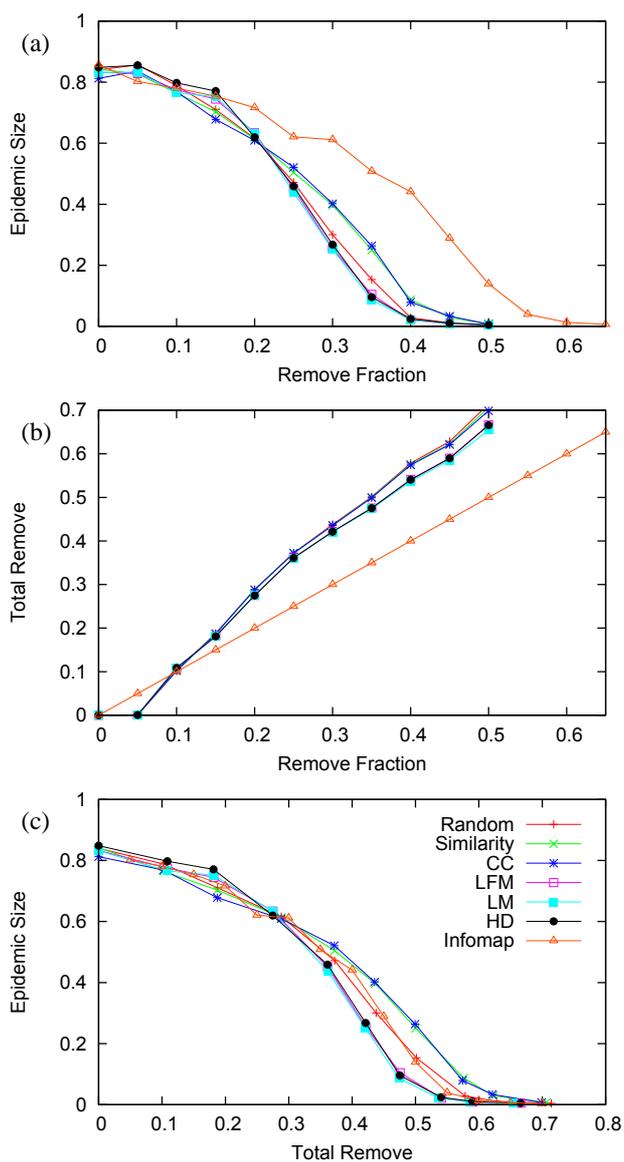

FIG. 2. (Color online) Effect of control strategies on SIR epidemic spreading in the random (Erdös-Rényi) network. Similarity, CC, LFM, LM, and HD are local strategies; Infomap is a global strategy. (a) Epidemic size, as fraction of network size, plotted against remove fraction. (b) Fraction of edges removed, plotted against remove fraction. (c) Epidemic size plotted against fraction of edges removed.

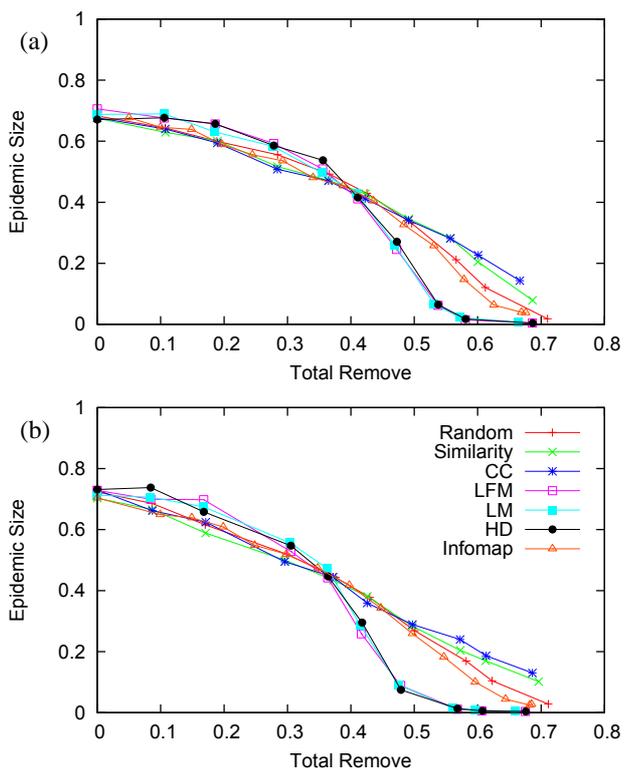

FIG. 3. (Color online) Effect of control strategies on SIR epidemic spreading in synthetic networks. Similarity, CC, LFM, LM, and HD are local strategies; Infomap is a global strategy. The plots show epidemic size as fraction of network size. (a) Exponential network. (b) Scale-free network.

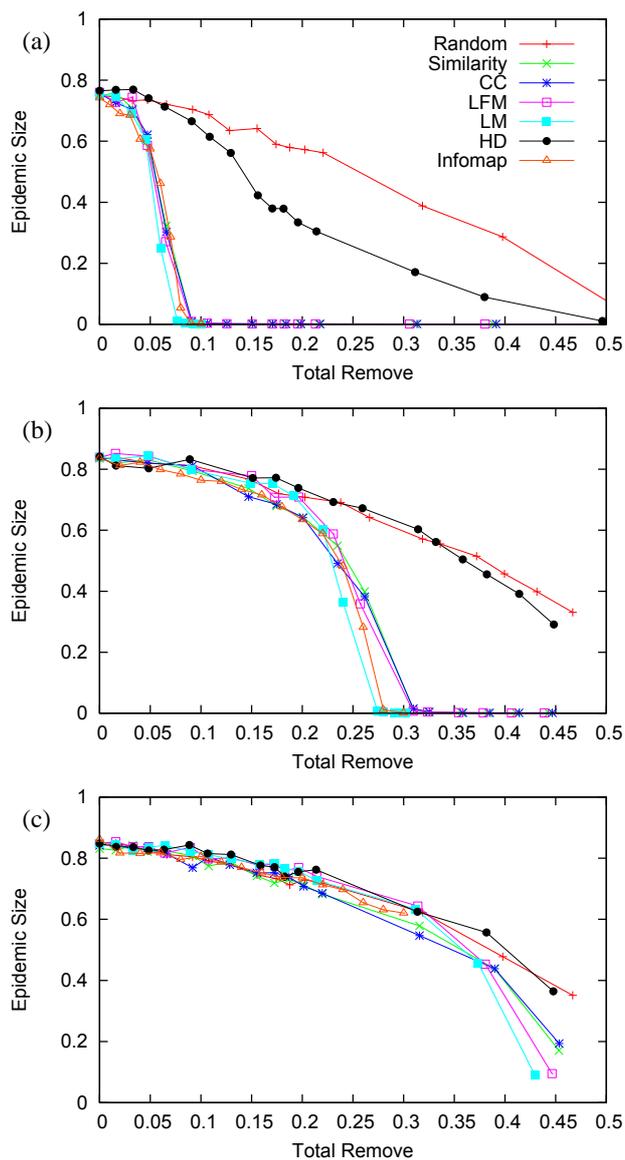

FIG. 4. (Color online) Effect of control strategies on SIR epidemic spreading in 10000-vertex LFR network. Similarity, CC, LFM, LM, and HD are local strategies; Infomap is a global strategy. The plots show epidemic size as fraction of network size. Each plot is for a different amount of mixing, to show the influence of the strength of community structure. (a) Mixing parameter 0.1. (b) 0.3. (c) 0.5.

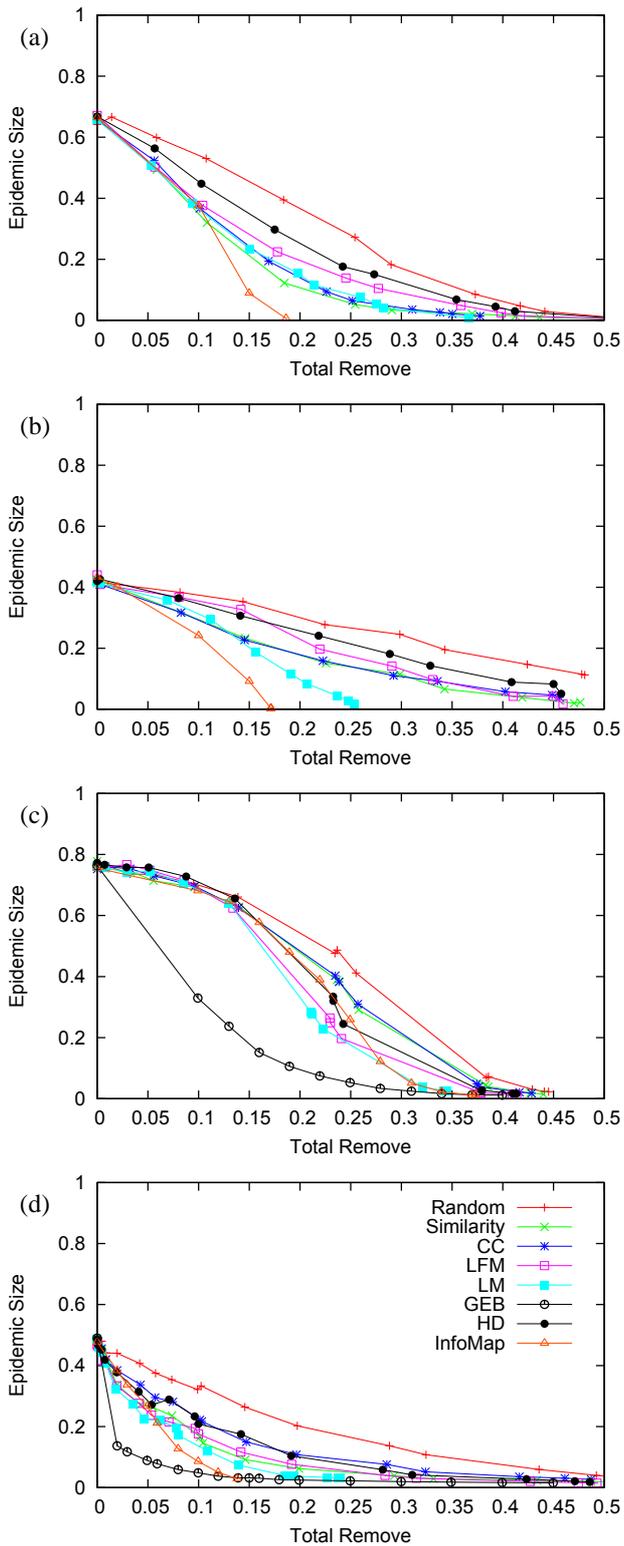

FIG. 5. (Color online) Effect of control strategies on SIR epidemic spreading in real-world networks. Similarity, CC, LFM, LM, and HD are local strategies; Infomap and GEB are global strategies. The plots show epidemic size as fraction of network size. (a) Blogs network. (b) PGP network. (c) School network. (d) Netscience network.

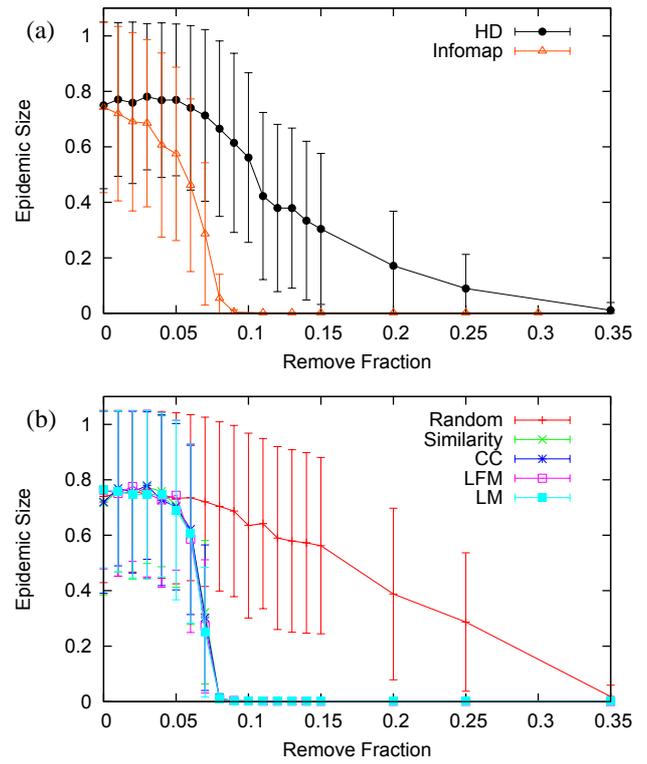

FIG. 6. (Color online) Effect of control strategies on SIR epidemic spreading in the 10000-vertex LFR network. Similarity, CC, LFM, LM, and HD are local strategies; Infomap is a global strategy. (a) Variation in epidemic size for Infomap and HD strategies. (b) Variation in epidemic size for other strategies.

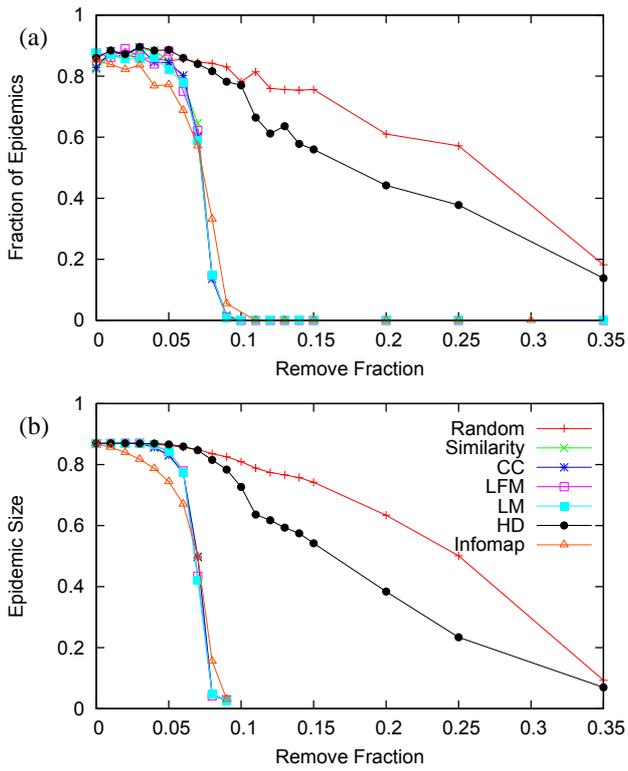
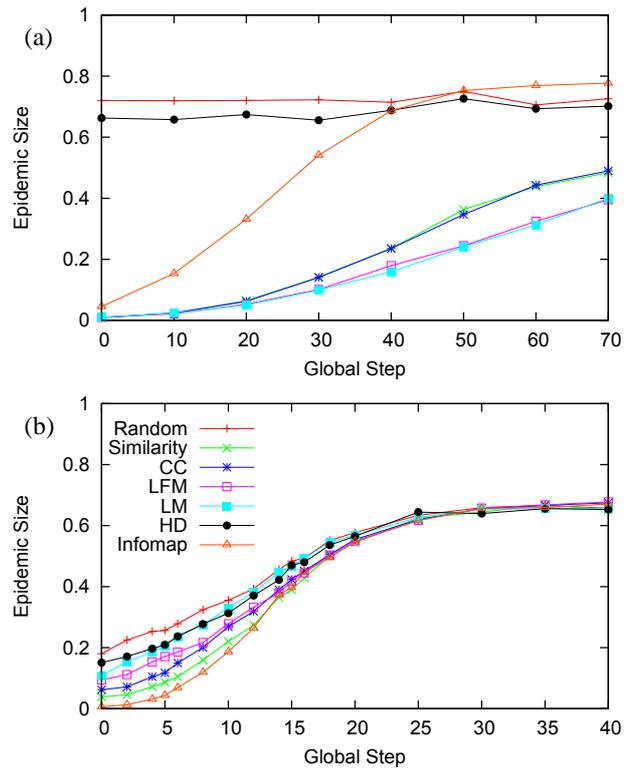

FIG. 7. (Color online) Effect of control strategies on SIR epidemic spreading in the 10000-vertex LFR network with mixing parameter 0.1. Similarity, CC, LFM, LM, and HD are local strategies; Infomap and GEB are global strategies. (a) Fraction of epidemics whose size is greater than 2% of the network size. (b) Average epidemic size of epidemics whose size is greater than 2% of the network size.

FIG. 9. (Color online) Effect of delay ("global step") before applying control strategies in the global belief-based model, with SIR epidemic. The plots show epidemic size as fraction of network size. (a) LFR network (remove fraction 20%). (b) Blogs network (remove fraction 30%). We do not show the fraction of edges removed; this is constant (about 0.1 for LFR and 0.2-0.3 for Blogs) because the strategies are not affected by time or disease prevalence.

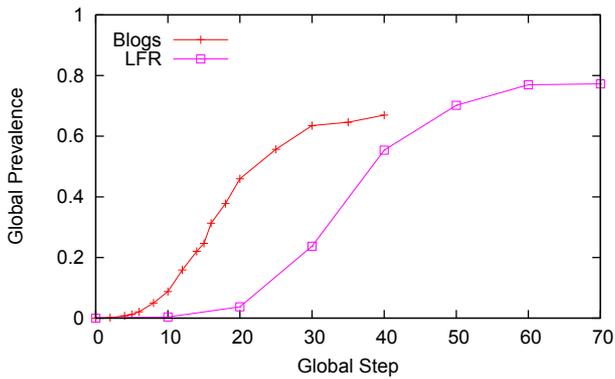

FIG. 8 (Color online) Variation of disease prevalence with time (named "global step") in the absence of control strategies, with an SIR epidemic. LFR network and Blogs network.

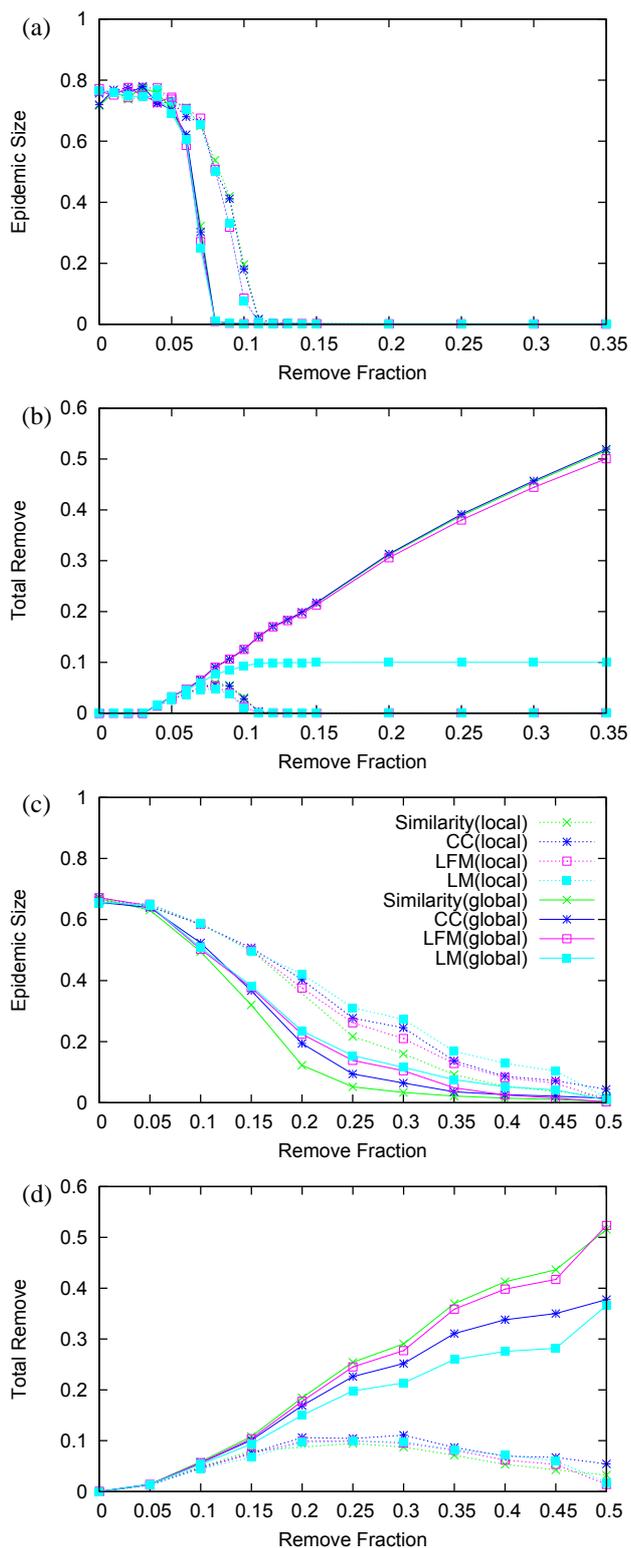
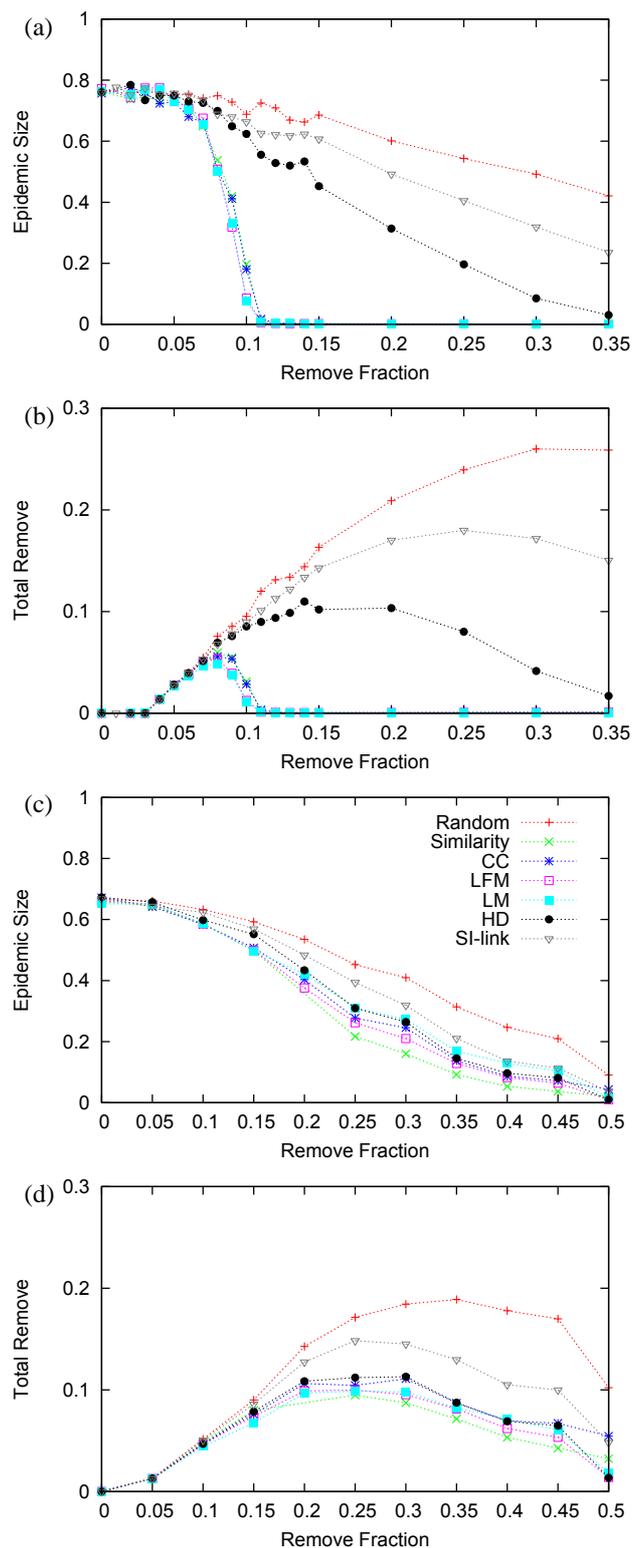

FIG. 10. (Color online) Comparison of global and local prevalence-based models using local control strategies, in an SIR epidemic. With the local model, the strategy is applied by a vertex when it becomes infected. With the global model, the strategy is applied in advance. (a) LFR network: epidemic size as fraction of network size. (b) LFR network: fraction of edges removed. (c) Blogs network: epidemic size. (d) Blogs network: edges removed.

FIG. 11. (Color online) Comparison of local prevalence-based model (the control strategy is applied by a vertex when it becomes infected) and the SI-link strategy, in an SIR epidemic. (a) LFR network: epidemic size as fraction of network size. (b) LFR network: fraction of edges removed. (c) Blogs network: epidemic size as fraction of network size. (d) Blogs network: fraction of edges removed.

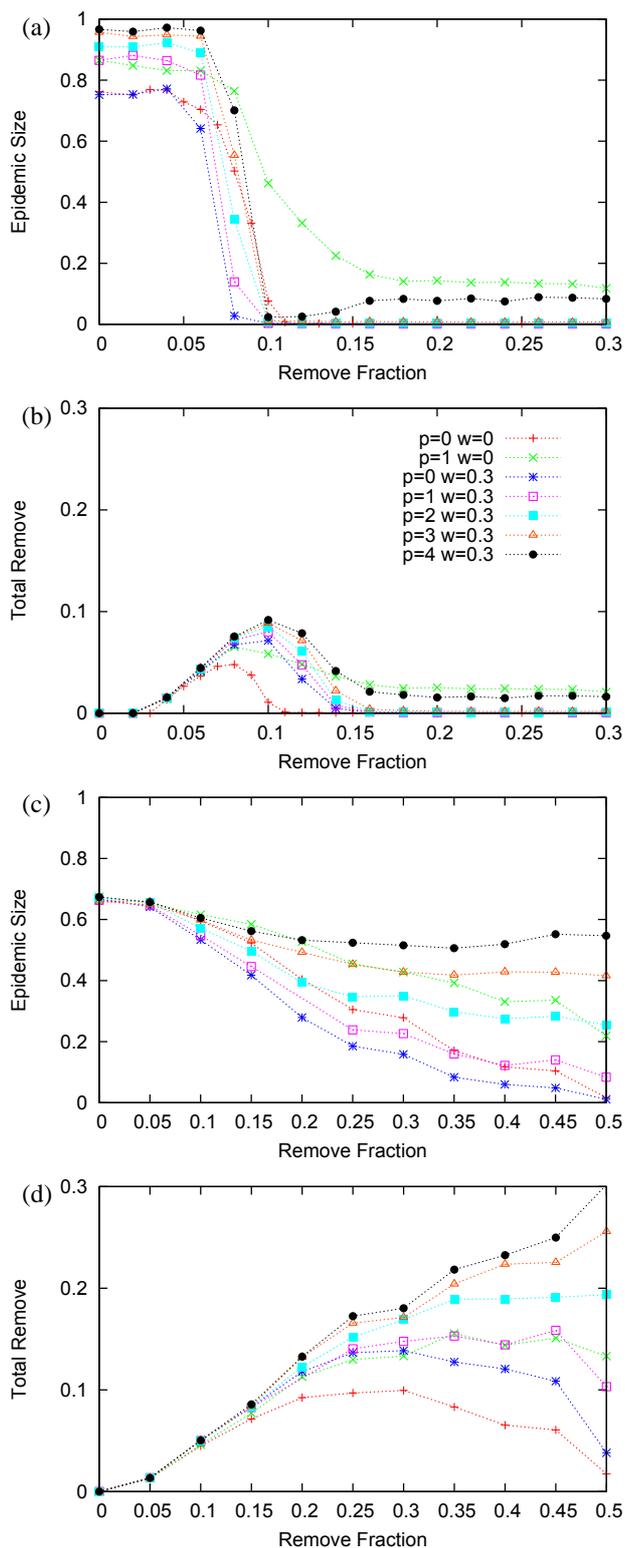

FIG. 12. (Color online) The effect of awareness spreading in a SPIR epidemic. The local belief-based model is used: the Local modularity strategy is applied by a vertex when it becomes aware. $p$ is the duration of the presymptomatic infectious phase. $w$ is the rate of awareness spreading. (a) LFR network: epidemic size as fraction of network size. (b) LFR network: fraction of edges removed. (c) Blogs network: epidemic size as fraction of network size. (d) Blogs network: fraction of edges removed.